\documentclass[3p, 11pt]{elsarticle}
\usepackage[utf8]{inputenc}
\usepackage{graphicx}
\usepackage{siunitx}
\biboptions{}
\usepackage{xcolor}
\usepackage{amsmath}
\usepackage{lineno}
\usepackage{setspace}
\begin{document}
\begin{frontmatter}
\title{Structure and hardness of \textit{in situ} synthesized nano-oxide strengthened CoCrFeNi high entropy alloy thin films}

\author[label1]{Subin Lee\corref{cor}}
\author[label2]{Dominique Chatain}
\author[label1]{Christian H. Liebscher}
\author[label1]{Gerhard Dehm\corref{cor}}
\address[label1]{Max-Planck-Institut für Eisenforschung GmbH, D-40237 Düsseldorf, Germany}
\address[label2]{Aix-Marseille Univ, CNRS, CINaM, 13288 Marseille, France}
\cortext[cor]{Corresponding author\\
E-mail address: s.lee@mpie.de (S. Lee), dehm@mpie.de (G. Dehm).}

\date{\today}

\begin{abstract}
    In this study, we report on face-centered cubic structured CoCrFeNi high-entropy alloy thin films with finely dispersed nano-oxide particles which are formed by internal oxidation. Analytical scanning transmission electron microscopy imaging found that the particles are $\mathrm{Cr_2 O_3}$. The oxide particles contribute to the hardening of the film increasing its hardness by 14\% compared to that of the film without precipitates, through the Orowan-type strengthening mechanism. Our novel approach paves the way to design medium- and high-entropy alloys with high strength by making use of oxide phases.

\end{abstract}

\end{frontmatter}


High entropy alloys (HEAs) have attracted tremendous interest in the last decades because of their impressive property combinations such as high strength and high ductility \cite{otto2013, gludovatzScience2014, miracle2017, yeh2004, zhang2014}. 
Initially, studies were focused on the alloy development and the characterization of equiatomic face-centered cubic (FCC) single-phase CoCrFeMnNi HEAs, the so-called Cantor alloy \cite{yeh2004, cantor2004}. 
However, it turned out that the strength and ductility of the Cantor alloy family at room and elevated temperatures are not sufficient enough to compete with conventional alloys, such as stainless steel or Ni-based superalloys \cite{zhang2014}.
Therefore, various traditional alloy design concepts have been applied to HEAs. For example, the introduction of metastable hexagonal close-packed phases increases the flow stress and promotes strain hardening dramatically through phase transformation during deformation \cite{linature2016}. Alloying with carbon \cite{wuAllComp2015, liActa2019} or boron \cite{seolActa2018} also enhances the yield strength and ultimate tensile strength of HEAs significantly.

One of the most common traditional alloying strategies to increase the strength is precipitation hardening: uniformly dispersed hard particles, such as oxide, nitride, or carbide particles, within a metal grain hinder dislocations motion and thereby strengthen it.
Since it has been reported that several additional phases become stable within the FCC HEAs on heat treatment \cite{schuh2015, pickering2016, ottoActa2016, he2017phase}, precipitation-hardened HEAs have been developed by tuning the alloy composition and/or controlling microstructure through thermomechanical processing \cite{he2016Acta, liu2016, liang2018NComm}.
For instance, minor additions of Ti and Al to a CoCrFeNi alloy increase yield strength from 200 MPa to 650 MPa with an outstanding 39\% elongation by forming $\mathrm{L}1_2$-structure precipitates \cite{he2016Acta}.
On the other hand, alloying of a CoCrFeNi HEA with Mo stabilizes $\sigma$ and $\mu$ phases increasing the hardness and strength of the alloy \cite{liu2016}.
Oxide-dispersion strengthening, which has been widely used in steels for elevated temperature applications \cite{UKAI2002749}, has been also applied to HEAs by using powder metallurgy \cite{moravcik2018, DOBES201899, chen2020MatDesign} and has produced a superior increase in strength and hardness.
Due to the inherent limitation of the sintering process, however, the size and spatial distribution of the oxide particles is not homogeneous and the interface between oxides and matrix is not strong enough to achieve outstanding elongation. 
As an example, plasma sintered CoCrFeNiTi HEA showed an ultimate tensile strength of 1460 MPa, whereas the ductility is limited to less than 15\% elongation to failure \cite{moravcik2018}.

Thin film structured HEAs have also been intensively studied \cite{dolique2009, liao2017, Addab2020Acta, li2018microstructures} not only for combinatorial screening \cite{marshal2019}, but also because of their excellent mechanical properties, such as hardness \cite{cheng2011}, wear and corrosion resistance \cite{cheng2013}.
Thin film deposition techniques can be used as a material processing tool to promote internal oxidation via rapid oxygen diffusion and to achieve additional strengthening from dispersed particles.

In the present study, we \textit{in situ} synthesized oxide nanoparticle-strengthened CoCrFeNi HEA thin films.
Through annealing of an ultrafine-grained sputter-deposited thin film with a columnar grain structure promoting fast oxygen diffusion, we obtained a uniform distribution of few tenths of nanometer-sized chromium oxide particles embedded in the HEA matrix. 
The microstructure of the annealed thin film and the oxide particles were analyzed by using high-resolution and analytical scanning transmission electron microscopy (STEM).
Finally, the impact of the oxide nanoparticles on the mechanical properties of the film was characterized by nanoindentation.

Thin films were deposited by magnetron co-sputtering of pure metal targets of Co (99.99 wt.\%, MaTecK), Cr (99.95 wt.\%, MaTecK), Fe (99.99 wt.\%, Evochem), and Ni (99.995 wt.\%, K.J. Lesker). The base pressure of the ultra-high vacuum chamber was $4.5 \times 10^{-4}$~\si{\pascal} and the Ar pressure during sputtering was $10^{-1}$~\si{\pascal}.
C-sapphire substrates were cleaned before the deposition with low power radio frequency Ar ion bombardment. 
A more detailed procedure for the thin film deposition can be found in our previous work \cite{Addab2020Acta}. The overall deposition rate was 0.13 \si{\nano\meter\per\second} at room temperature and the final thickness of the film was 500 \si{\nano\meter}. 
Then, the wafer-film sample was annealed for one hour in a different vacuum chamber at 1273 \si{\kelvin} (or 0.74 of the melting temperature of the film) under a vacuum pressure of $10^{-4}$~\si{\pascal}.

The phases and texture of the films were analyzed by X-ray diffraction (XRD) using Co-K$\alpha$ radiation in a Seifert diffractometer equipped with a high voltage generator, a poly-capillary beam optic, and a scintillation detector. 
Scanning electron microscopy (SEM) imaging and electron backscatter diffraction (EBSD) analysis were carried out in a focused ion beam (FIB)-SEM dual-beam workstation (Auriga, Zeiss).
STEM specimens were prepared by a site-specific plan-view FIB lift-out \cite{field2004} using a FIB-SEM dual beam workstation (Scios 2, Thermo Fisher Scientific). 
STEM images were acquired using an aberration-corrected STEM operated at 300 kV (Titan Themis 60-300, Thermo Fisher Scientific). The probe current was about 80 pA. The collection angles for the high-angle annular dark-field (HAADF) images were set to 78–200 mrad using a semi-convergence angle of 23.8 mrad. Scan noise and specimen drift during the acquisition was compensated by rigid registration of image series of about 50 frames acquired with a dwell time of 1~\si{\micro\second}. 
For energy-dispersive X-ray spectroscopy (EDS) the probe current was raised to 100 pA and X-rays were collected with the ChemiSTEM system (Thermo Fisher Scientific). The total acquisition time was 25 min. All EDS quantifications were performed using K peaks.
The energy resolution determined from the full-width at half maximum of the zero energy loss peak was 0.9 eV for the electron energy loss spectroscopy (EELS, GIF Quantum ERS, Gatan). Dual-EELS mode was used to increase the signal-to-noise ratio at the high-loss region.

Nanoindentation tests were conducted after removing the oxide scale and surface pores by gentle mechanical polishing by using 0.01 \si{\micro\meter} diamond lapping film. 
The nanoindentation was performed with a Hysitron G200 using a Berkovich indenter tip under load-controlled mode.
The maximum load was set to 500 \si{\micro\newton}.


\begin{figure*}[t]
	\centering
	\includegraphics[width=0.9\textwidth]{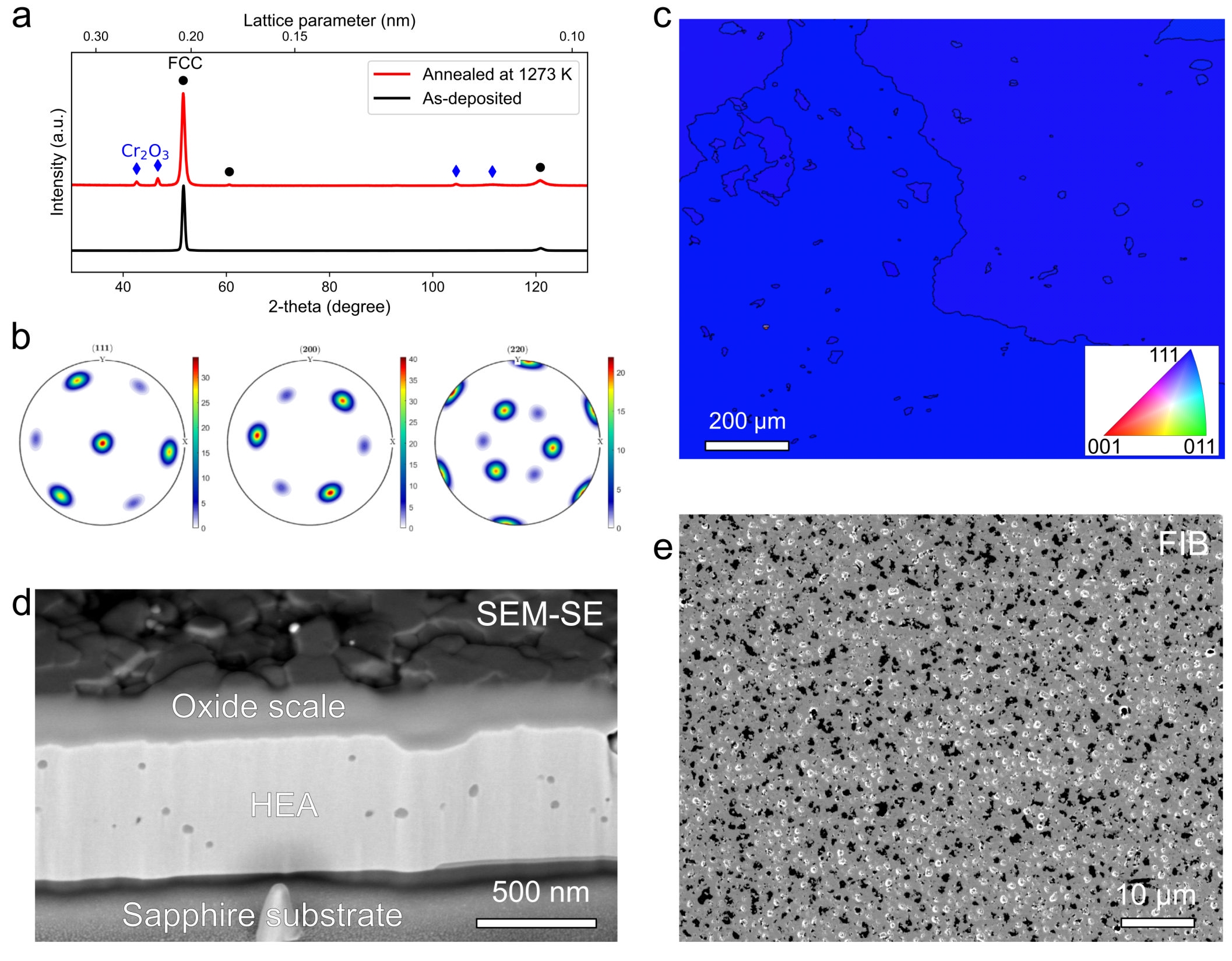}
    \caption{
    The microstructure of as-deposited and anznealed thin films. (a) XRD $\theta/2\theta$ scans before and after heat treatment. Peaks marked with black circles indicate an FCC phase with a lattice parameter of 0.356 nm. After annealing additional peaks appear which correspond to the $\mathrm{Cr_2 O_3}$ phase marked by blue diamonds. (b) XRD texture analysis on the as-deposited film shows a strong orientation relationship to the c-sapphire substrate. (c) An EBSD orientation map from the annealed film. All grains keep a (111) texture while their size increases to a few hundred micrometers. (d) FIB cross-section image of the film after annealing. (e) Top-view SEM image of the film. Dark contrasts are remaining oxide scales and brighter contrasts are from surface pores.
    }
    \label{fig:microstructure}
\end{figure*}
The XRD phase analysis shows that the as-deposited CoCrFeNi thin film is FCC single phase with a strong (111) texture as only (111) and (222) peaks are visible (Fig. \ref{fig:microstructure}a). 
The shape of the spectrum remains similar even after one hour of annealing at 1273 \si{\kelvin} except the additional peaks which correspond to $\mathrm{Cr_2 O_3}$ phase from the surface oxide scale as marked by blue diamonds.
Strong 3-fold (111) peaks in XRD pole figure maps of the as-deposited film in Fig. \ref{fig:microstructure}b suggest that the grains have a preferred in-plane orientation relationship with the c-sapphire substrate.
From EBSD pole figure analysis (Supplementary Figure 1), a specific orientation relationship is found, which is referred to as OR2 \cite{Addab2020Acta, dehm2002growth, curiotto2011}.
It can be written as:
\begin{equation}
	\mathrm{HEA}(111)\pm[\bar1{1}0] \; \; ||  \;\; \alpha\mathrm{-Al_2 O_3}(0001)[11\bar{2}0]
\end{equation}
here $\pm$ indicates two variants in twin relationship. 
Weaker reflections with 60\si{\degree} rotation in the pole figures are from the twin-oriented grains.
The EBSD inverse pole figure map from the annealed film shows the (111)-oriented texture (Fig. \ref{fig:microstructure}c). 
The grain size of the as-deposited film was a few tens of nanometers measured by FIB imaging, and has increased to several hundred micrometers after annealing except small islands inside the large grains.
The cross-section of the annealed film was observed by FIB-SEM, which shows not only a thick $\mathrm{Cr_{2}O_{3}}$ scale on top of the film but also oxide particles inside the film (darker contrast inside the film in Fig. \ref{fig:microstructure}d).
After removing the surface scale by Ar plasma etching, the surface morphology was analyzed by ion channeling contrast imaging (Fig. \ref{fig:microstructure}e). 
Bright features are surface pores and the dark contrast is stemming from the remaining chromium oxide scale on top of the pores. 
The average size of surface pores is in the range of 100 nm. 

\begin{figure*}[t]
	\centering
	\includegraphics[width=0.9\textwidth]{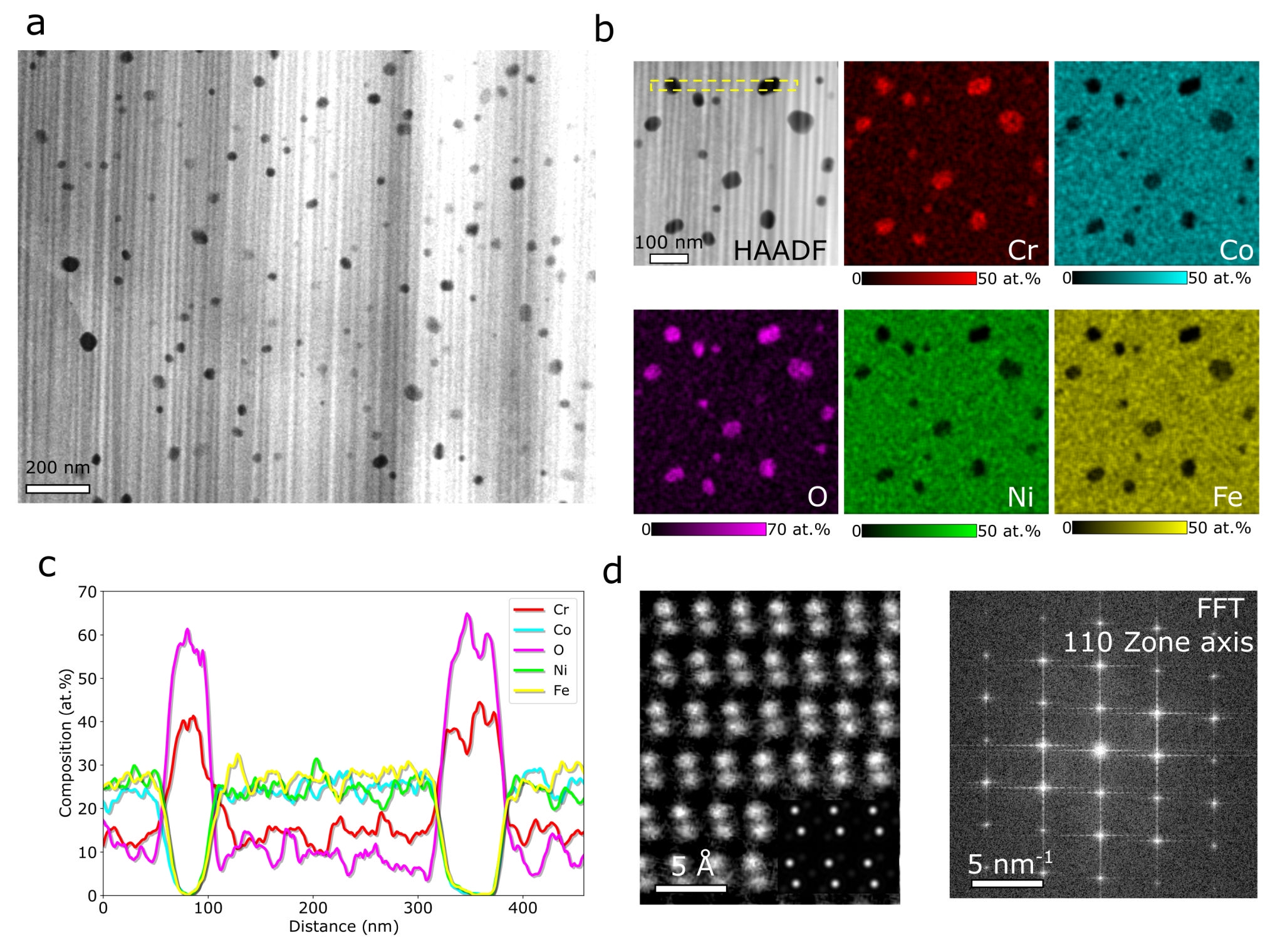}
    \caption{
    Microstructure analysis of the annealed film by STEM. (a)STEM-HAADF image showing the distribution of $\mathrm{Cr_2 O_3}$ nanoparticles. (b) STEM-EDS elemental maps representing the high concentration of Cr and O in the particles. (c) Intensity profile of elemental maps which is averaged over the yellow box in (b). Cr and O ratio is close to 3:2 and the concentration of other elements in the particles is negligible. (d) High-resolution STEM-HAADF and fast Fourier transformed images of the oxide particles in the (110) zone axis. }
    \label{fig:eds}
\end{figure*}

The internal microstructure of the annealed film was analyzed by STEM imaging with a plan-view lift-out sample (Fig. \ref{fig:eds}a). 
Dark particles are uniformly distributed through the entire sample with an average size and areal density of $12.7 \pm 7.0$ nm (here $\pm$ shows the standard deviation of measurement) and 4.5\% of the total area, respectively. 
STEM-EDS shows that the stoichiometry of the particles is close to $\mathrm{Cr_2 O_3}$ while the composition of the matrix is $\mathrm{Co_{27}Cr_{13}Fe_{31}Ni_{27}}$ [at.\%], Cr depleted with respect to the equiatomic composition (Figs. \ref{fig:eds}b and c).
 Inside the particles, the X-ray signals from other elements (Co, Fe, Ni) are negligible as shown in the line profiles extracted from the yellow box in Fig. \ref{fig:eds}b. It suggests that those particles are pure $\mathrm{Cr_2 O_3}$ without forming solid-solution. 
High-resolution HAADF imaging (Fig. \ref{fig:eds}d) also matches well with simulated images (inset in Fig. \ref{fig:eds}d) using an underlying trigonal (R-3C) space group of the $\mathrm{Cr_2 O_3}$ crystal structure. 
Fast Fourier transform of the high-resolution STEM image also confirmed the trigonal crystal structure of the particles.
Most of the interfaces between the particles and the matrix are incoherent except the basal plane of the particles which is semi-coherent with the (111) planes of the matrix with a lattice misfit strain of $\sim6\%$ measured by STEM-HAADF imaging.

\begin{figure}[t]
	\centering
    \includegraphics[width=0.5\textwidth]{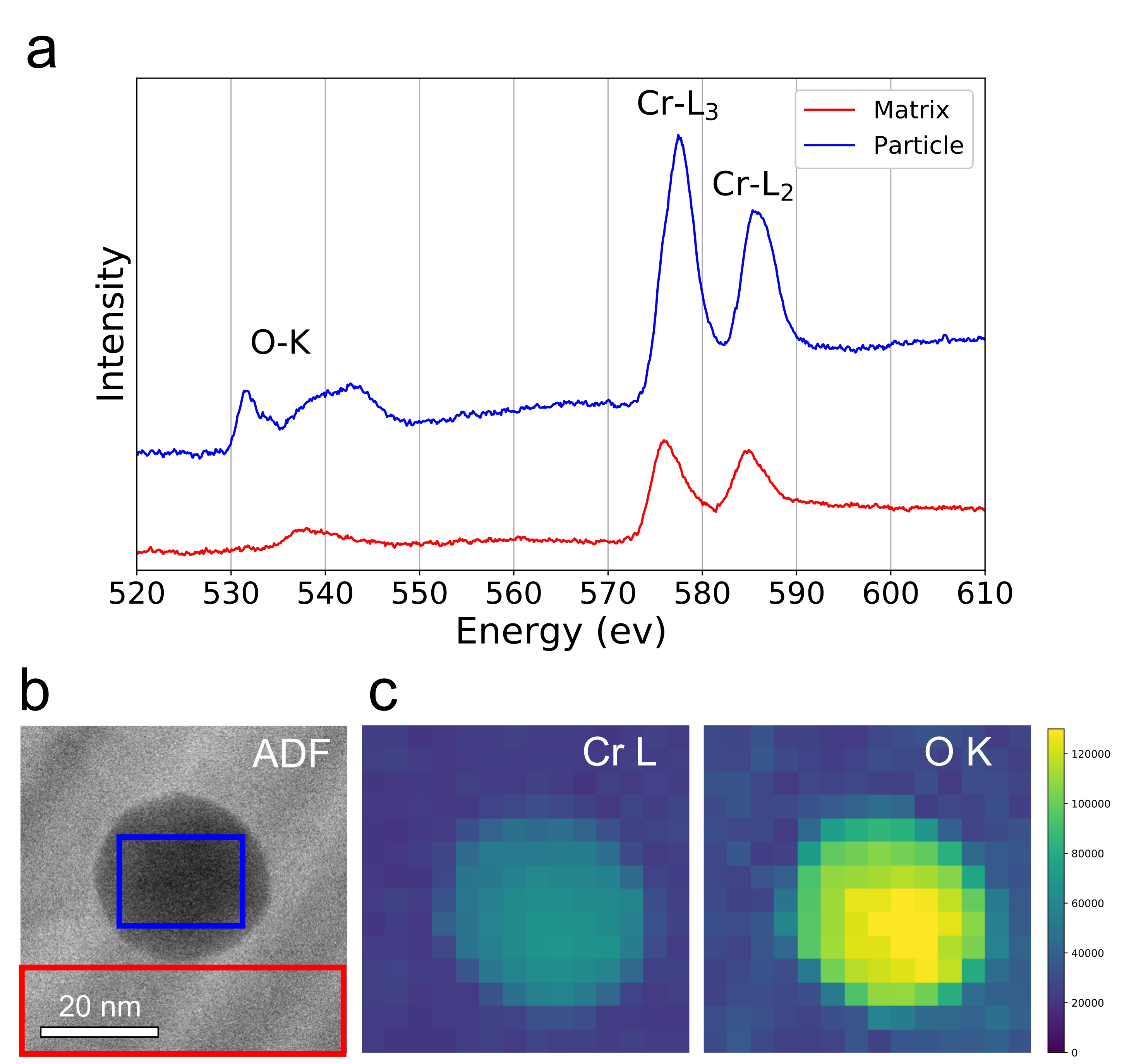}
    \caption{
    STEM EELS analysis on a $\mathrm{Cr_2 O_3}$ particle. (a) Comparison of the spectrum; one is averaged from the particle region (blue box in (b)) and the other is averaged from the matrix (red box in (b)). (b) STEM-ADF image of the $\mathrm{Cr_2 O_3}$ particle (c) EELS intensity maps at Cr-L edge and O-K edge.
    }
    \label{fig:EELS}
\end{figure}

Due to the limited energy resolution of $\sim$130 eV of EDS, it is challenging to distinguish the Cr-L (573 eV) and O-K peaks (523 eV) in an EDS spectrum, which can introduce errors in the quantitative EDS analysis.
Therefore, we further analyzed the local bonding environment of Cr and O by STEM-EELS mapping. 
The spectra shown in Fig. \ref{fig:EELS}a are averaged over the red and blue boxes in Fig. \ref{fig:EELS}b, respectively. 
There are two distinct differences between the two spectra. 
First, the spectrum from the oxide particle (blue) clearly shows a double peak in the O-K edge fine structure, whose shape matches well with that previously reported for $\mathrm{Cr_2 O_3}$ \cite{suzuki1997}.
The spectrum extracted from the matrix has a small hump at 535 eV which is most likely stemming from the surface oxide on the TEM specimen. 
Second, the intensity ratio of the Cr-L$_{2}$ and -L$_{3}$ peaks is different in the matrix and the particle since Cr is in a different oxidation state in the oxide compared to the metallic bonding in the FCC lattice structure \cite{castillo2008, daulton2006}.
In summary, the careful phase analysis of the particles by high-resolution STEM-imaging, -EDS, and -EELS reveal that nano-sized $\mathrm{Cr_2 O_3}$ particles form in the HEA thin film upon annealing.

We obtained a FCC CoCrFeNi thin film with finely distributed oxide particles within the grains, while it has been reported that oxide scales are only found at the surface of annealed bulk Cantor alloy family.
The internal oxidation could derive from the low Cr content of the film because the oxygen permeability and the size of the internal oxidation zone are inversely proportional to Cr contents in Fe-Cr alloys \cite{prillieux2017, setiawan2010}.
Also, the initial nanocrystalline microstructure with a grain size of a few tenths of nanometers could provide additional oxygen to the matrix by trapping oxygen contamination during deposition and/or oxygen diffusion along grain boundaries.

\begin{figure}[t]
	\centering
	\includegraphics[width=0.5\textwidth]{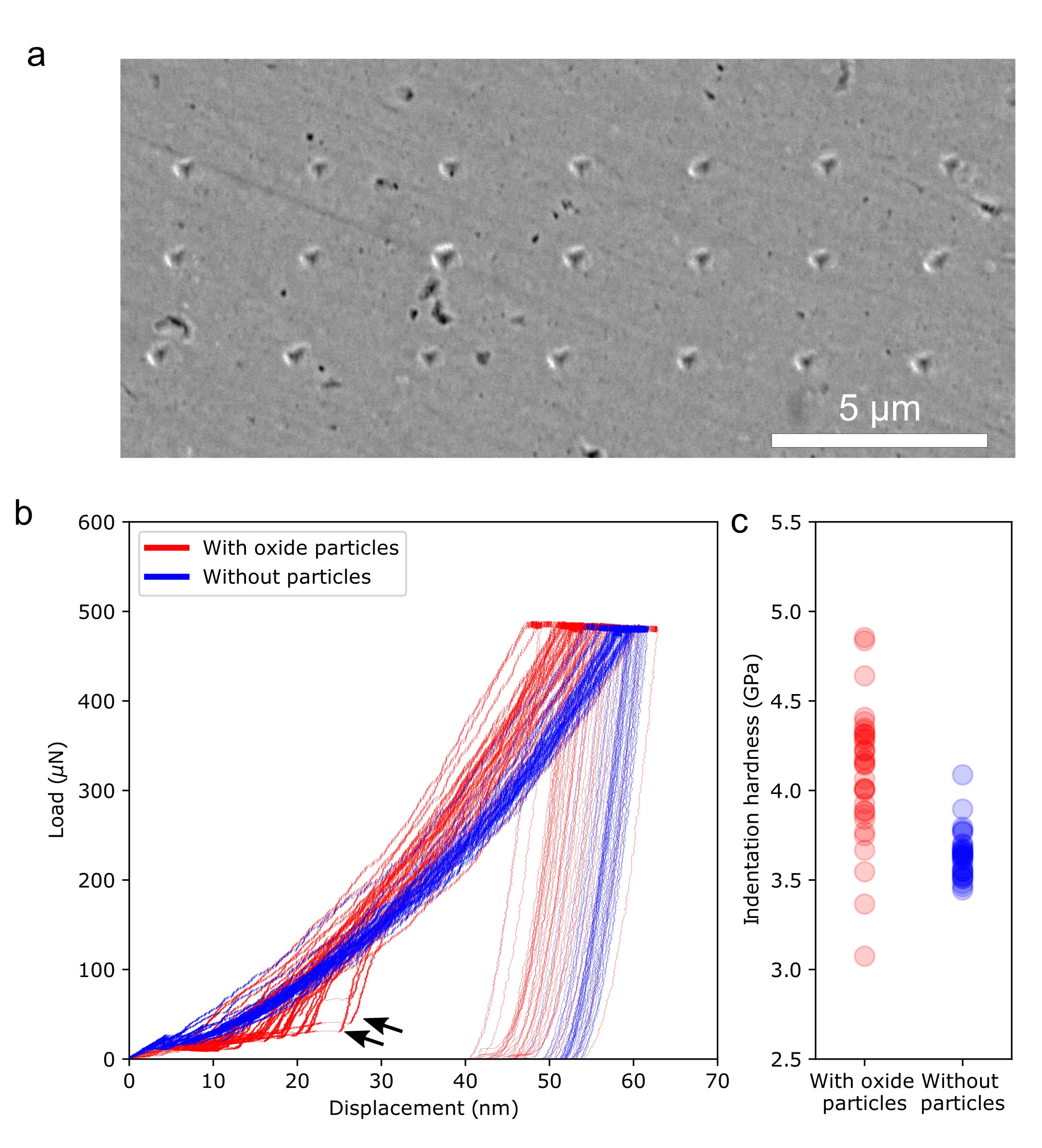}
    \caption{
    Comparison of the mechanical properties of the thin film with a HEA film without oxide particles. (a) A SEM image of the film surface after indentation. The indentation data which was obtained from surface imperfections, such as remaining oxide particles or surface scratches are excluded for further data processing. (b) Load-displacement curves from two samples. 30 indentations were made in each sample. (c) The hardness and modulus of the two films show that $\mathrm{Cr_2 O_3}$ particles strengthen the film increasing hardness 15\%. 
    The arrows indicate pop-in in the film with Chromium oxide precipitates.}
	\label{fig:indentation}
\end{figure}

Lastly, we characterized the effects of the finely dispersed $\mathrm{Cr_2 O_3}$ nanoparticles on the hardness of the thin film by using nanoindentation.
A (111)-textured equiatomic CoCrFeNi thin film was used as a reference sample that does not contain any oxide particles and has a large grain size to exclude the influence of grain boundaries.
Fig. \ref{fig:indentation}a shows imprints on the film surface after nanoindentation. The load-displacement data where imprints are located at surface imperfections, such as remaining oxide particles or surface scratches, are excluded for further data analysis.
Load-displacement curves of the oxide-containing film show a higher fluctuation and the obtained hardness values also exhibit a larger deviation compared to the reference sample, which is most likely originating from the high density of oxide particles (Fig. \ref{fig:indentation}b).
The mean hardness value of the film with and without particles is $4.1 \pm 0.36$ GPa and $3.6 \pm 0.13$ GPa, respectively (Fig. \ref{fig:indentation}c).
The oxide-containing film shows distinct pop-in behaviors at the beginning of indentation (arrows in Fig. \ref{fig:indentation}b), which might be related to imperfections on the surface of the film.


Strengthening mechanisms in polycrystalline materials can be categorized into four groups: solid-solution hardening, grain boundary hardening, Taylor hardening by dislocations, and precipitation hardening.
In our case, the effects of grain boundaries can be ignored because we tested on the interior of the large grains of the film.
Dislocation strengthening, which is proportional to the square root of the dislocation density, should be also negligible since the dislocation density was less than $10^{12}$ \si{\per\meter\squared} after annealing measured by TEM investigation of underneath the indents. 
Therefore, the dominating strengthening mechanisms are hardening by the dispersed oxides and the inherent solid solution hardening of the HEA matrix, in other words, the lattice friction stress which is higher than in conventional alloys \cite{lee2020, utt2020}. 

There are two types of particle strengthening mechanisms active depending on particle geometry, distribution, and their mechanical strength: 1) Orowan-type dislocation bowing and 2) dislocations cutting through particles.
Orowan strengthening dominates over the latter when the particles are impenetrable for dislocations, incoherent with the matrix, or their size is large.
Based on the fact that the dispersed oxide particles are incoherent and have a higher hardness (30 GPa \cite{kao1989}) compared to that of the HEA matrix (2-3 GPa \cite{bracq2019}), we expect that the Orowan mechanism is dominant since the particles do not deform with the matrix and the active stress is necessary to bow dislocations expanding from particle to particle.
The contribution from particle strengthening following the Ashby-Orowan mechanism can be determined as following \cite{arzt1998size, kamikawa2015}:
\begin{equation}
	\sigma_{Ashby-Orowan} = \dfrac{0.8 m G b}{2 \pi \sqrt{1-\nu} L} \ln\left(\dfrac{x}{2b}\right)
\end{equation}
where $m$ is Schmid factor (0.272 is used considering (111) film normal orientation), $G$ is shear modulus (= 84 GPa \cite{wu2014temperature}), $b$ is the magnitude of Burgers vector, and $\nu$ is the Poisson's ratio (=0.28 \cite{wu2014temperature}). $L$ is the average inter-particle spacing, $L = \sqrt{2/3}  \left( \sqrt{\pi/f} -2 \right) r$ and $x$ is the average particle diameter, $x = 2\sqrt{2/3} r$, where $r$ is the radius of the particles. $f$ is the volume fraction of the particles, which is 1.5\%, which is calculated from the projected areal fraction of the precipitates (4.5\%) in a STEM-HAADF image. A stereological correction was done by assuming the thickness of the TEM sample as 50 nm \cite{sonderegger2006}. 
Using the equation above, the contribution of the particles on the strength of the film is estimated to increase of 240 MPa of yield strength or 720 MPa of hardness while the difference in hardness in our experimental results is 500 MPa. 

One of the possibilities for the discrepancy is that the size, shape, and distribution of the particles are not ideal while the Orowan strengthening model assumes a uniform distribution of same-sized spherical particles.
In our case, the radius varies from a few nm to 30 nm and the shape of particles are far from the ideal spherical shape. 
Furthermore, the shallow maximum indentation depth may play an important role which causes indentation size effects \cite{oliver1992improved}.
Since the film thickness is around 500 nm, we had to limit the indentation depth to 50 nm, 10\% of the sample thickness, to minimize the effects of the substrates.
However, as the size of the indent becomes smaller, the hardness value we measure is affected by indentation size effects and dislocations operate under high shear stress. 
Therefore, the Orowan strengthening mechanism is not as effective as in the ideal case as the dislocations may penetrate small particles.

In summary, we synthesized nano oxide-dispersion strengthened HEA thin film and characterized its structure and hardness by advanced analytical STEM techniques and nanoindentation. The key conclusions are summarized as follows:

\begin{itemize}
    \item By annealing (111) textured CoCrFeNi thin film at 1273 \si{\kelvin} for 1 hour under $10^{-4}$~\si{\pascal} vacuum, we obtained hundreds of micrometers large-grained thin film with homogeneously distributed fine $\mathrm{Cr_2 O_3}$ particles.
    
    \item The structure and composition of the oxide particles are analyzed by STEM imaging, EDS, and EELS, and all experiments indicate $\mathrm{Cr_2 O_3}$.
    
    \item The average particle radius is $12.7 \pm 7.0$ nm, and the volume fraction of precipitates is 1.5\% of the total volume.
    
    \item The oxide particle-containing film shows a 14\% higher hardness compared to the one without particles.
    It is speculated that the dominating hardening mechanism is Orowan-type dislocation bowing.
    
\end{itemize}

\section*{Acknowledgement}
This work was supported by the DFG-ANR collaborative project ”Analysis of the stability of High Entropy Alloys by Dewetting of thin films” under the grant DE796/11-1 (MPIE) and ANR-AHEAD-16-CE92-0015-01 (CINaM); SL acknowledges support by the Alexander von Humboldt Foundation (MPIE).  
 
\bibliographystyle{elsarticle_num_no_title}
\bibliography{reference.bib}

\end{document}